\documentclass{aa}
\usepackage{graphics}
\begin{document}

   \thesaurus{06     
              (11.09.1 M74;      
               11.09.1 NGC613;   
               11.09.1 NGC7741;  
               11.19.2;          
               11.19.5)}         

   \title{ The nature of near-infrared emission from
    spiral galaxies}

   \subtitle{}

   \author{P.A. James
          \inst{1}
          \and
          M.S. Seigar\inst{2}}

   \offprints{P. James}

   \institute{ Astrophysics Research Institute, 
   Liverpool John Moores University,
   Twelve Quays House, Egerton Wharf, Birkenhead CH41~1LD, UK\\
              email: paj@astro.livjm.ac.uk
         \and
             Department of Mathematical Physics \& Astronomy,
             Universiteit Gent, 
             Krijgslaan 281 (S9), B-9000 Gent,
             Belgium\\
             email: Marc.Seigar@rug.ac.be
             }

   \date{}
   \titlerunning{Near-IR emission from spiral galaxies}

   \maketitle

   \begin{abstract}
We present K-band spectroscopy for several regions of three nearby
spiral galaxies, NGC~613, NGC~628 and NGC~7741.  Analysis of the depth
of the 2.293$\mu$m CO absorption feature in these spectra reveals that
some regions have deep absorptions, indicative of recent star
formation, while others have anomalously shallow absorptions. We
interpret the latter as evidence for a significant but localised
contribution to the 2.3~$\mu$m light from hot dust with an effective
temperature of $\sim$1000~K, which could have a significant effect on
the K-band morphologies of star-forming galaxies.

      \keywords{ galaxies: spiral - 
                 galaxies: stellar content}
   \end{abstract}

%

\section{Introduction}
The stars which dominate the stellar mass in galaxies are effectively
invisible.  Most of the optical light in the solar neighbourhood comes
from A dwarfs and K \& M giants, whereas the local stellar mass is
dominated by G, K and (particularly) M dwarfs, which contribute a
negligible fraction of the visible light (Kroupa, Tout \& Gilmore
\cite{krou}, Binney \& Merrifield \cite{binn}).  In actively star-forming
systems the disparity between the luminous and mass-bearing stellar
components is even greater.  For this reason, and because of the
diminished effects of dust obscuration, near-infrared (near-IR)
imaging has become a key observational technique for determining the
morphologies and stellar contents of galaxies.  For example, near-IR
imaging has been widely used to determine the underlying structure of
spiral arms in disk galaxies (e.g. Rix \& Rieke \cite{rix93}, Block et
al. \cite{bloc}, Rix \& Zaritsky \cite{rix95}, Seigar \& James
\cite{seib}), since arms are both extremely dusty and associated with
star formation. In this paper we will look in more detail at the
assumption that near-IR light is a good tracer of the stellar mass in
galaxies.

The hot OB stars which can dominate the blue light of a star-forming
galaxy have relatively little impact on the K-band (2.2~$\mu$m)
emission (Persson et al. \cite{pers}).  However, by $\sim$10$^7$~years after
the onset of a burst of star formation, the most massive stars will
have evolved to the red supergiant phase and will be contributing
significantly to the near-IR light output (Persson et al. \cite{pers}).  In
addition, hot (600--1000K) dust associated with OB associations gives
rise to mid-infrared emission with a short-wavelength `tail' which
extends into the K window (Doyon, Joseph \& Wright \cite{doyo}).  For
extreme `starbursting' objects, the supergiant plus dust emission can
exceed that of the old stellar population in the K-band (e.g. Ridgway,
Wynn-Williams \& Becklin \cite{ridg}, Doyon et al. \cite{doyo},
Puxley, Doyon \& Ward \cite{puxl}).  For normal disk galaxies, the
effects will be more modest, but still significant.

Rhoads (\cite{rhoa}) has addressed the question of the supergiant
contribution to K-band disk light by mapping the depth of the
2.3~$\mu$m CO photospheric absorption over the disks of three nearby
spiral galaxies.  This feature is stronger in supergiant stars than in
the old stellar population (Frogel et al. \cite{frog}, Doyon et al.
\cite{doyo}), and extends over a sufficiently broad wavelength range
to be easily quantified as a colour change in narrow band filters.
Using this technique, Rhoads (\cite{rhoa}) determines the supergiant
contribution to the three galaxies (NGC~278, NGC~2649 and NGC~5713) to
be only $\sim$3$\%$ of the overall near-IR flux, but rising locally in
regions of strong star formation to as much as 33$\%$.  This is an
important result, if generally true, since this is a significant
fraction of the arm-interarm contrast in near-IR images of normal
spirals (Rix \& Zaritsky \cite{rix95}, Seigar \& James
\cite{seib}), making it very uncertain whether near-IR features
reflect changes in the overall stellar mass density, or in the
location of the young supergiant population.  Given an age of
1--2$\times$10$^{7}$ years, and assuming a formation velocity
dispersion of $\sim$10~kms$^{-1}$ for Population I stars, the
supergiants will have typically moved only 0.1--0.2~kpc from their
birthplaces in the disk, whereas the late K or early M giants which
dominate the red light from the older stellar population (Frogel et al. 
\cite{frog}, James \& Mobasher \cite{jame}) will be well mixed with
other stellar types. Thus it is important to quantify the relative
contributions from giant and supergiant stars.

Doyon et al. (\cite{doyo}) point out that extreme caution must be used in
interpreting filter CO measurements of the type used by Rhoads
(\cite{rhoa})
in regions of strong star formation, due to hot dust, non-thermal
emission and extinction, which can significantly affect the shape of
the underlying galaxy continuum.  More reliable measurements can be
obtained using long slit spectroscopy, with the wavelength range
covering both the CO absorption and the continuum emission at shorter
wavelengths, which enables the continuum slope to be determined
explicitly.  This is the strategy we have adopted for the present
paper.  We have obtained low-resolution, long-slit spectroscopy
covering the full K-band for bulge and disk locations in NGC~613 and
NGC~628, and for the bar of NGC~7741.  We present spectroscopic CO
indices for all these regions, and use them to determine the fraction
of light contributed by supergiant stars.  We also compare the
spectroscopic properties of bulges with those of inner disk regions,
and with elliptical galaxies.

It should be noted that Oliva et al. \cite{oliv} conclude that CO
absorption strength is not a clear indicator of recent star formation
in stellar populations.  They find that an old population of
high-metallicity giants can have the same CO index as a starburst
dominated by low metallicity supergiants.  However, their data (see
Table 2 of Oliva et al. \cite{oliv}) do show significantly deeper CO
absorptions in both the centres of HII galaxies and young (6--17~Myr)
LMC clusters than in 4 comparison ellipticals or the bulges of 3
spiral galaxies.  Thus, whilst there is clearly scope for ambiguity of
interpretation, and the possibility of metallicity effects must be
considered when interpreting results, empirically it appears that CO
strength does contain information on star formation history.

The organisation of the present paper is as follows. Section 2
contains a description of the galaxies observed, and of the
spectroscopic observations obtained.  The data reduction procedures
are described in section 3, and section 4 contains the interpretation
of the observed spectral indices in terms of the fraction of light
from old and young stars, and from hot dust, in the different galaxy
regions.  Section 5 summarises the conclusions reached, and contains
suggestions for future work to be undertaken in this area.

\section{Observations}

\subsection{Instrumental set-up}

All observations presented here were taken on four half-nights,
1997 August 3,5,6 and 7.  The instrument used was the facility near-IR
spectrometer CGS4 on the United Kingdom Infrared Telescope, UKIRT.
This was configured with the short, 150~mm focal length camera, giving
1\farcs23 pixels, and the 75 line/mm grating.  The grating
angle was set to give a total spectral range of 1.97--2.63$\mu$m
(larger than the atmospheric K window), and the dispersion was
0.00258$\mu$m/pixel.  We used the 2--pixel wide slit, oriented N--S
for all observations, and moved the array by one pixel between
sub-integrations to give automatic bad pixel replacement, and Nyquist
sampling over the resolution permitted by the slit.  The slit is
90$^{\prime\prime}$ long. For NGC~7741 this was large enough for us to
keep the galaxy bar on the slit in both beamswitch positions, but the
larger sizes of NGC~613 and NGC~628 required the observation of
completely offset positions for sky monitoring.  This was done by
alternating one-minute integrations at the main and sky positions.

At regular intervals, A-type stars were observed to permit variations
in atmospheric transparency to be ratioed out of the galaxy spectra.
A-stars were chosen as they are much too hot to have molecules in
their photospheres, and thus they have no CO absorptions (and indeed
no other strong features around the wavelength range of interest
here). Care was taken to observe stars at the same airmass as the
corresponding galaxy.

\subsection{ Observed galaxies}

NGC~613 is a barred spiral, classified as SBbc in the Third Reference
Catalogue of Bright Galaxies (de Vaucouleurs et al. \cite{deva};
henceforth RC3).  It has a major axis diameter of 5\farcm5, a total
blue magnitude of $B_T=$10.73, a recession velocity of 1475~kms$^{-1}$,
and a distance of 17.9~Mpc (Jungwiert, Combes \& Axon \cite{jung}),
assuming H$_0=$75~km~s$^{-1}$Mpc$^{-1}$ (throughout the present paper)
and correcting for a Virgo infall model.  Thus 1$^{\prime\prime}$
corresponds to 86~pc at NGC~613.  It shows clear sign of an active
galactic nucleus, and V\'eron-Cetty \& V\'eron (\cite{vero})
classified it as a `Composite' galaxy, with central spectral features
showing both Seyfert activity and vigorous star formation.

NGC~628 (M74) is classified as an Sc (RC3), and has a major axis
diameter of 10\farcm5 and a total blue magnitude of $B_T=$9.95.
Sharina, Karachentsev \& Tikhonov (\cite{shar}) calculate a distance of
7.3~Mpc for NGC~628, such that 1$^{\prime\prime}$ corresponds to
35~pc.  It has a recession velocity of 657~kms$^{-1}$.  There is no evidence of
any nuclear activity in NGC~628 (AGN or vigorous star formation), and
Wakker \& Adler (\cite{wakk}) note that the density of molecular
material has a local minimum around the nuclear region.  They find
evidence for a ring of enhanced CO emission with a diameter of
$\sim$45$^{\prime\prime}$, centred approximately on the nucleus.

Finally, NGC~7741 is a strongly barred spiral, classified as SBcd by
RC3.  It has a recession velocity of 751~kms$^{-1}$, a diameter of
4\farcm4 and a total blue magnitude of $B_T=$11.84.  Tully
(\cite{tull}) gives its distance as 12.3~Mpc, such that 1$^{\prime\prime}$
projects to 60~pc.

\section{Data reduction}

All data reduction was performed using the FIGARO and KAPPA packages
provided by the Starlink project.  Bias subtraction, flat fielding,
bad pixel removal and sky subtraction were done
automatically at the telescope.  Some sky line residuals remained, and
these were removed by fitting across regions of sky at the ends of the
slit.  One-dimensional spectra were then extracted by adding rows of
pixels clearly containing galaxy signal.  Simple addition of these
rows was found to be more reliable than using the FIGARO `optimal
extraction' procedure.  The A-star spectra were extracted in the same
way, and were then divided into each of the galaxy spectra to correct
for spectral variation in the atmospheric transmission.  No attempt
was made to flux calibrate any of the spectra.

The next stage was to correct for variations in the level and slope of
the continuum in the galaxy spectra.  This was done using a routine
written by Dr Clive Davenhall under the Starlink QUICK facility.  The
routine fits a power-law function to continuum regions of the
spectrum, with the only free parameters being the index of the power
law and a multiplicative normalisation constant (see Doyon et al. 1994
for a physical justification of this choice of function). Regions
affected by strong spectral features (principally Brackett $\gamma$
2.166$\mu$m, and of course the CO absorption) are masked and not
included in the fit.  The routine produces a pure power-law `spectrum'
over the full wavelength range of the input spectrum.  The latter is
then divided by the former to produce a `rectified' spectrum with
a continuum level of 1.0 at all wavelengths.

These rectified spectra were then wavelength calibrated using CGS4
argon arcs taken near in time to the corresponding spectra, and the
wavelength scales were corrected to the galaxy rest frame using the
recession velocities given above.

It is simple to measure CO absorption strength from a rectified
spectrum, simply by calculating the mean level over some standard
wavelength interval.  In the present paper, we use two measures.  The
first is the spectroscopic index $CO_{sp}$ proposed by Doyon et al.
(\cite{doyo}):
$$
CO_{sp} = -2.5 log <R_{2.36}>
$$
where $<R_{2.36}>$ is the mean level in a rectified spectrum between
2.31 and 2.4$\mu$m.  The second measure was proposed by Puxley et al. 
(\cite{puxl}) to give strong discrimination between different
stellar populations, and is the equivalent width (henceforth $EW$) in nm of
the CO absorption between 2.2931 and 2.32$\mu$m.

\begin{figure}[h]
\includegraphics{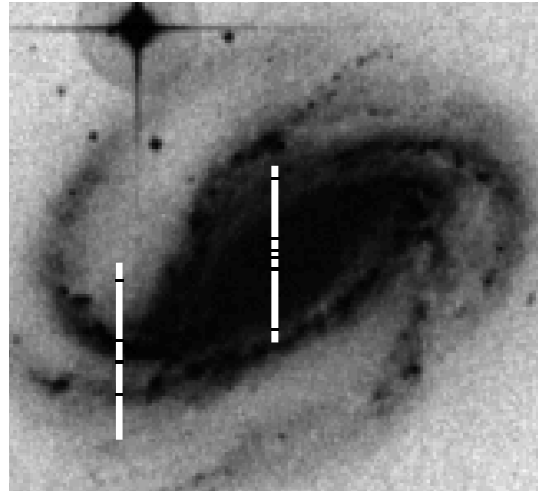}
\vspace*{7.0cm}
\caption{Digitised Sky Survey image of NGC~613 showing the slit
positions used.  Black lines on the slit show the limits of the 
slit regions used for spectral extraction (see text for full
explanation). North is up and east to the left; the scale is given by
the 90$^{\prime\prime}$ long slit.}
\label{fig1}
\end{figure}

\begin{figure}[h]
\includegraphics{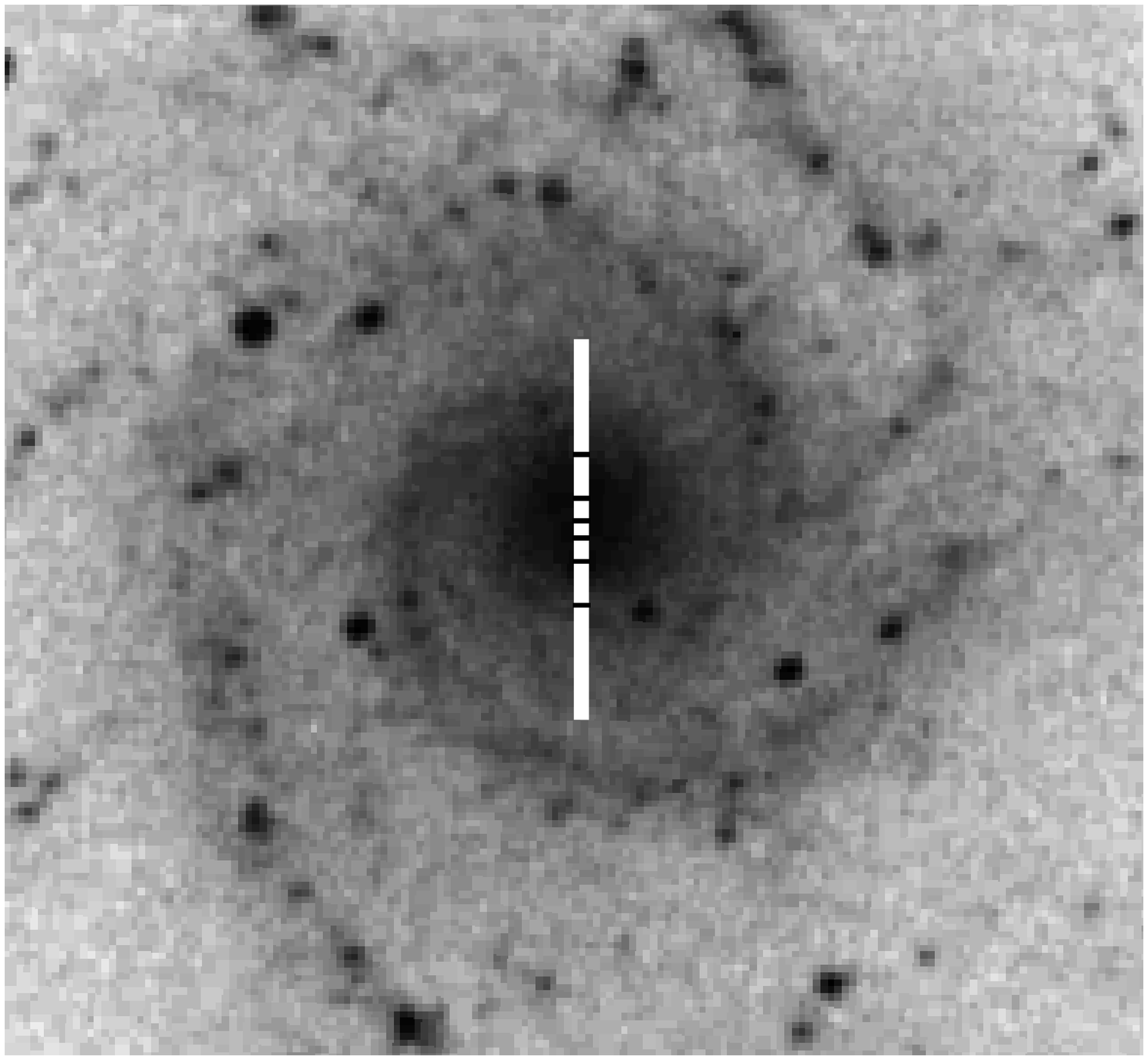}
\vspace*{7.0cm}
\caption{Digitised Sky Survey image of NGC~628 showing the slit
position used.  Black lines on the slit show the limits of the three
slit regions used for spectral extraction (see text for full
explanation).}
\label{fig2}
\end{figure}

\begin{figure}[h]
\includegraphics{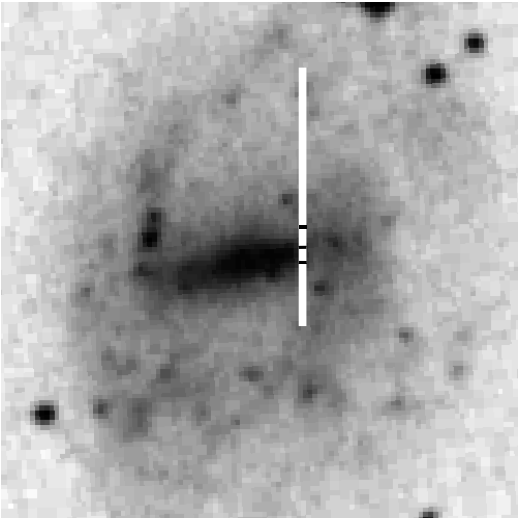}
\vspace*{7.0cm}
\caption{Digitised Sky Survey image of NGC~7741 showing the slit
position used.  Black lines on the slit show the limits of the two
slit regions used for spectral extraction (see text for full
explanation).}
\label{fig3}
\end{figure}

Table 1 contains $CO_{sp}$ and $EW$ values for the following regions of the
target spirals:

{\bf NGC~613}: The 2\farcs5 wide slit was first positioned
on the infrared emission peak at the centre of the galaxy, in a N--S
orientation (this position angle was used for all spectra discussed
here).  Spectra were extracted for three regions along the slit: the
4$^{\prime\prime}$ length centred on this nuclear position (Region 1),
the 16$^{\prime\prime}$ centred on the nucleus (Region 2), and an
outer bulge/inner disk region sampling light from
8$^{\prime\prime}$--39$^{\prime\prime}$ from the nucleus.  The slit
was also positioned on a strong spiral arm, 80$^{\prime\prime}$ W and
50$^{\prime\prime}$ S of the nucleus. Again, three spectra were
extracted from this slit position.  Region 4 comprised the
11$^{\prime\prime}$ centred on the K-band peak from the arm, Region 5
was the 17$^{\prime\prime}$ immediately S of Region 4 and Region 6 the
31$^{\prime\prime}$ immediately N of Region 4.  These slit positions
are shown superimposed on an optical image of NGC~613 in
Fig.~\ref{fig1}, and the horizontal black lines on the slits delimit
the positions of Regions 1--6.  Note that the slit positions shown in
Figs. 1--3 are only approximate, since the slits were actually centred
on the K-band peak of the galaxy, which may not be coincident with the
optical centre of the galaxy.

{\bf NGC~628}: The slit was positioned on the
infrared emission peak at the centre of the galaxy.  Spectra were
extracted for three regions along the slit: the 4$^{\prime\prime}$
centred on the nucleus (Region 1), the 15$^{\prime\prime}$
centred on the nucleus (Region 2), and an outer bulge region
sampling light 7$^{\prime\prime}$--18$^{\prime\prime}$ from the
nucleus (Region 3).  The slit position and region delimiters are shown
in Fig.~\ref{fig2}.

{\bf NGC~7741}: The slit was positioned 15$^{\prime\prime}$ W of the
nuclear infrared peak, and lay almost perpendicular to the bar.  The spectra
were noisy, and only two regions could be usefully defined: Region 1
was the 5$^{\prime\prime}$ centred on the K-band bar peak, and Region
2 the 7$^{\prime\prime}$ immediately N of Region 1. The slit position
and region delimiters are shown in Fig.~\ref{fig3}.

\begin{figure}[h]
\includegraphics{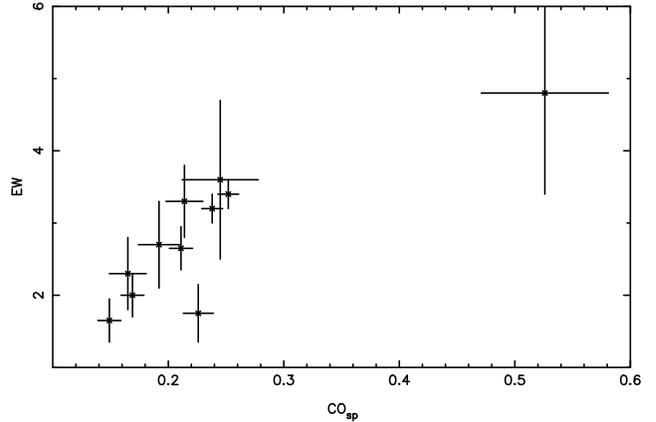}
\vspace*{7.0cm}
\caption{CO equivalent width $EW$ plotted against CO spectral index
CO$_{sp}$ for the 11 regions listed in Table 1.}
\label{fig4}
\end{figure}

Fig.~\ref{fig4} shows the correlation between the two measures of CO
strength, CO$_{sp}$ and $EW$, where the data points are taken from
Table~\ref{tab1}.  Most of the points show a good correlation between
the two estimates, consistent with the estimated errors.  These errors
included two components. The first was calculated from the standard 
deviation in the fitted continuum points, on the assumption that the
noise level remains constant through the CO
absorption, giving an error on both the continuum level and on the
mean level in the CO absorption, whch were added in quadrature.  The
second error component came from the formal error provided by the
continuum fitting procedure.  This procedure could leave a residual
tilt or curvature in the spectrum, and the formal error was used to
quantify this contribution.

Two of the points in fig.~\ref{fig4} (both from the disk of NGC~613)
are displaced significantly to the right of this correlation,
indicating anomalously low $EW$ values or (more probably) anomalously
high CO$_{sp}$ values. This latter could occur due to errors in the
extrapolating of the underlying continuum from 2.28~$\mu$m to
2.4~$\mu$m, and may indicate that the second error component mentioned
above was underestimated, at least for the CO$_{sp}$ values.  This is
likely to be a problem for rapidly star-forming systems, where
continuum slope variations are more likely, and argues for using the
$EW$ definition of Puxley et al. \cite{puxl} for such objects.

\section{Analysis of CO indices}

\begin{figure}[h]
\includegraphics{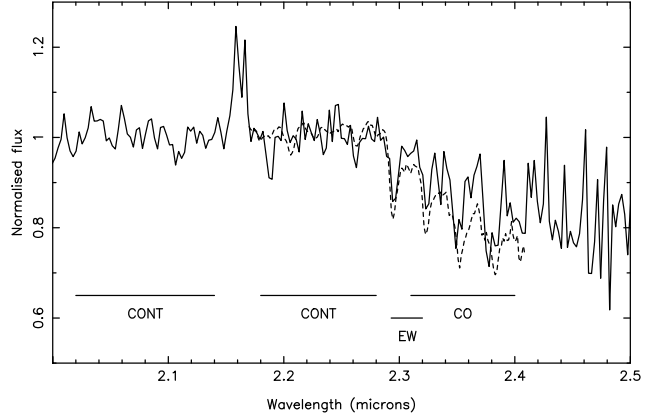}
\vspace*{7.0cm}
\caption{Spectrum of NGC~7741 Region 1 (solid line) compared with the
mean elliptical galaxy spectrum from James \& Mobasher (1999) (dashed
line)}
\label{fig5}
\end{figure}

\begin{figure}[h]
\includegraphics{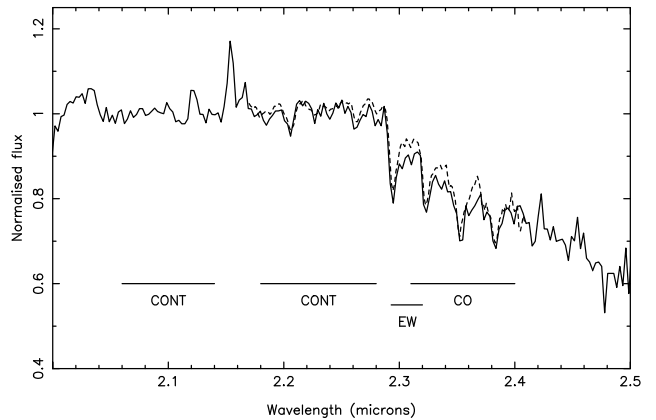}
\vspace*{7.0cm}
\caption{Spectrum of NGC~613 Region 1 (solid line) compared with the
mean elliptical galaxy spectrum from James \& Mobasher (1999) (dashed
line)}
\label{fig6}
\end{figure}

We will now discuss the CO strengths measured in the 3 spiral
galaxies.  It is useful to compare these with the results of James \&
Mobasher (\cite{jame}) who made CO absorption measurements, with the
same telescope/instrument combination, for a sample of 50 elliptical
galaxies from rich cluster, group and isolated field environments.
They found that the weakest CO absorptions occurred in the 10 isolated
elliptical galaxies, which have $<CO_{sp}> =$0.23$\pm$0.01, $<EW>
=$2.8$\pm$0.1~nm.  We take these as the baseline levels for a pure old
stellar population, but with relatively high metallicity, and the mean
spectrum for the 10 ellipticals is shown for comparison purposes as
the solid line in Fig.~\ref{fig5} and Fig.~\ref{fig6}. (Note that a
different instrumental set-up was used for the elliptical spectroscopy
and the wavelength range is correspondingly smaller.) CO strengths
were typically found by James \& Mobasher (\cite{jame}) to be larger
by 10--15\% for cluster ellipticals, presumably due to metallicity and
star formation history effects.  For a young stellar population,
1--2$\times$10$^7$~years after a burst of star formation, population
synthesis models predict peak CO absorption strengths of $CO_{sp}
=$0.4$\pm$0.04, $EW =$5.3$\pm$0.5~nm (e.g. Rhoads
\cite{rhoa}, Origlia et al. \cite{orig}), which is supported by observations
of young LMC clusters by Persson et al. \cite{pers}.
We can then interpret any measured CO strengths lying between these
extremes as a weighted sum of young and old populations.

      \begin{table}
      \caption[]{Measured CO indices and $EW$ values}
         \label{tab1}
      \[
         \begin{array}{lccc}
            \hline
            \noalign{\smallskip}
            Galaxy  &  Region & CO_{sp} & EW (nm) \\
            \noalign{\smallskip}
            \hline
            \noalign{\smallskip}
$NGC$~613    &  1  &  0.252\pm0.009 & 3.4\pm0.2 \\
$NGC$~613    &  2  &  0.238\pm0.009 & 3.2\pm0.2\\
$NGC$~613    &  3  &  0.192\pm0.018 & 2.7\pm0.6\\
$NGC$~613    &  4  &  0.165\pm0.016 & 2.3\pm0.5\\
$NGC$~613    &  5  &  0.245\pm0.033 & 3.6\pm1.1\\
$NGC$~613    &  6  &  0.526\pm0.055 & 4.8\pm1.4\\
$NGC$~628    &  1  &  0.169\pm0.010 & 2.0\pm0.3\\
$NGC$~628    &  2  &  0.211\pm0.010 & 2.65\pm0.3\\
$NGC$~628    &  3  &  0.214\pm0.016 & 3.3\pm0.5\\
$NGC$~7741   &  1  &  0.149\pm0.010 & 1.65\pm0.3\\
$NGC$~7741   &  2  &  0.226\pm0.013 & 1.75\pm0.4\\
            \noalign{\smallskip}
            \hline
         \end{array}
      \]
   \end{table}

However, a surprising result shown in Table 1 is that several regions
show CO absorptions significantly {\em weaker} than those expected for
old stellar populations, regardless of whether $CO_{sp}$ or $EW$ is
used. This is the case for the arm (Region 4) of NGC~613, the central
nucleus (Region 1) of NGC~628, and the bar (Region 1) of NGC~7741
(Fig.~\ref{fig5}). Such an effect could in principle result from incorrect sky
subtraction, which could effectively cause a positive flux offset at
all wavelengths and hence decrease the measured $EW$ and CO$_{sp}$
values.  However, in every case there are spectra of
lower-surface-brightness regions extracted from the same observations,
which should show this effect even more strongly, but do not. 

The shallow CO features could be explained in terms of
metallicity effects.  Doyon et al. \cite{doyo} find a trend in CO
strength with metallicity such that 
$$\Delta CO_{sp}=0.11\Delta\left[Fe/H\right]$$ and Origlia et al.
\cite{ori97} find that low metallicity ($\left[Fe/H\right]$ between
--0.64 and --2.01) globular clusters have significantly lower CO
equivalent widths than do elliptical or spiral galaxies.  Thus
metallicity effects can clearly cause significant differences in CO
indices between different objects, or between different stellar
components (e.g. bulge, disk and halo) of a given galaxy.  However, we
do not favour this explanation for the CO index differences found in
the present paper, as this would require significantly lower
metallicities within, for example, the NGC~613 arm or the NGC~7741
bar, than in the disk material immediately surrounding these features,
and at the same galactocentric distances.

Low CO strengths could also be explained by extremely young starburst
populations, where even the near-IR light could be dominated by Main
Sequence stars.  However, supergiants cause a deep CO absorption in the
integrated light by $\sim$10$^7$~years after a burst of star formation
(Persson et al. \cite{pers}, Doyon et al. \cite{doyo}, Rhoads
\cite{rhoa}).  Thus it is hard to contrive a significant stellar
contribution to the light output which gives a CO depth as weak as
that measured in these three regions.

A final possibility is that there is a
significant non-stellar continuum at these wavelengths.  If present,
such a component would dilute the stellar light and reduce the
measured strength of the CO feature.  The most obvious possibility is
hot dust, which can radiate significantly in the K-band in strongly
star-forming regions of galaxies (Giuricin et al. \cite{giur}) and would be
expected to affect only localised regions within a given galaxy.  We
can put a lower limit on the fraction of the galaxy light at 2.3$\mu$m
contributed by such a component, by assuming that the underlying
stellar population must have a CO$_{sp}$ index of at least 0.23, and
an $EW$ of at least 2.8~nm. This gives two estimates of the fraction
of light which must be contributed by any dust emission in each of the three
regions of low indices.  For the NGC~628 nucleus, these fractions are
25\% and 27\% of the total light at 2.3$\mu$m, for the NGC~613 arm
27\% and 19\%, and for the NGC~7741 bar 34\% and 41\% respectively.
For any reasonable dust temperature (i.e. $<$1000K) such a component
would have a negligible emission at J and H, and hence J--K and H--K
colours should be redder by $\sim$0.3 magnitudes in such regions of
dust emission.  Spiral galaxy nuclei frequently show such colour
excesses (Glass \& Moorwood \cite{glas}, Seigar \& James \cite{seia}), but such
colour changes in arms (e.g. Rix \& Rieke \cite{rix93} ) and bars are
generally attributed to reddening by dust lanes.  

A cleaner test for dust emission would be to detect the emission at
longer wavelengths, e.g. in the L (3.8$\mu$m) band.  If
1000K dust contributes 20\% of the 2.3$\mu$m light, it should
contribute $\sim$40\% at 3.8 $\mu$m, assuming the stellar light to
have an effective temperature of 4000~K (K5 giants) and a
$\lambda^{-1}$ dust emissivity law.  For cooler dust the predicted L
excess is greater, and 600~K dust would have to be completely dominant
at 3.8$\mu$m to contribute significantly at 2.3$\mu$m.  Whilst the
latter is ruled out by measured JHKL colours of normal galaxies, K-L
colour excesses of 0.4--0.7~mag, as predicted for 1000K dust, are
frequently seen in star-forming nuclei (Glass \& Moorwood \cite{glas},
Giuricin et al. \cite{giur}).  It would be very interesting to test whether
such excesses are ever seen in strong bars or arms, as the hot dust
interpretation of our present data would seem to require.  Of course,
if there is a significant supergiant population in these regions, as
is very likely, the intrinsic CO depth would be greater than we have
assumed here, and the required dust component would be correspondingly
brighter.  With the present data it is impossible to disentangle these
two effects.

Table 1 also lists some regions with CO absorptions significantly
deeper than would be expected for an old stellar population. From the
measured $CO_{sp}$ indices and $EW$ values it is possible to estimate
the fraction of light contributed by old and young stellar
populations, but it should be emphasised that these are only
indicative estimates, and for these regions we neglect the effect of
any hot dust or other non-stellar continuum emission.  The region with
the most significant evidence for a young stellar population is the
central nucleus of NGC~613 (Region 1), shown in Fig.~\ref{fig6}, which
requires 13--24\% of the K-band light to be provided by a young
population. For Region 2 the fraction is down to 5--19\%, while for
Region 3 the CO strength is consistent with a purely old stellar
population.  Thus the outer bulge appears indistinguishable from an
elliptical galaxy (as was found by Oliva et al. \cite{oliv} for the 3
spirals in their sample), and while the central star formation found by
V\'eron-Cetty \& V\'eron (\cite{vero}) is clearly detected here, it
appears to be confined to the central $\sim$8$^{\prime\prime}$ in
radius.

NGC~613 Region 5 is a disk region south of a bright spiral arm, and
lies on its leading edge if we assume the arms to be trailing. This
region also has a deep CO absorption, implying that 10--30\% of the K
light comes from recently formed supergiants.  There is some evidence
for even deeper CO absorption in the disk region (6) on the trailing,
concave, side of the arm, but this spectrum is very noisy.

The final region showing marginal evidence for a supergiant population
is NGC~628 Region 3, the outer bulge, where $EW$ is
slightly larger than would be expected for an old stellar
population, and also larger than in the nuclear regions
of NGC~628.  Region 3 lies just within the peak of the ring of
molecular material detected in CO emission by Wakker \& Adler
\cite{wakk}, strengthening the identification of deeper CO absorptions
with recent star formation.

\section{Conclusions and future work}

The principal new result of this paper is that we have found regions
of two of the three disk galaxies studied where CO indices are
anomalously weak, even when compared with elliptical galaxies
specifically selected for the weakness of this feature. Whilst this
feature does show metallicity dependence, we feel that the required
variations in metallicity within the disks of the individual galaxies
studied weigh against this as the dominant cause of our findings. We
suggest that our observations indicate the presence of significant
continuum emission at 2.3~$\mu$m, most likely from dust at an
effective temperature of $\sim$1000K. This component appears to
contribute at least 20\% of the K-band light in some parts of these
galaxies, and would be clearly detected in mid-infrared imaging.

For the NGC~613 central bulge and disk interarm regions, and for the
outer bulge of NGC~628, we find evidence for supergiant emission
contributing 5--30\% of the local K-band surface brightness of these
galaxies.  These numbers are in good agreement with those measured by
Rhoads (\cite{rhoa}), and confirm his conclusion that caution should
be used when interpreting K-band imaging as an indicator of the old
stellar mass distribution in star-forming galaxies.  If there were a
significant hot dust component in these regions, the intrinsic CO
depths would have to be correspondingly greater, and hence the
fraction of supergiant light quoted above should be considered a lower
limit.

The outer bulge of NGC~613 has a K-band spectrum essentially identical
to that of the comparison elliptical galaxies, and thus appears to be
dominated by an old stellar population, subject to the uncertainties
given above.

We have thus identified a source of uncertainty in the interpretation
of K-band imaging of star-forming galaxies, in addition to the
supergiant light contribution studied by Rhoads (\cite{rhoa}). Some of
our spectra, and particularly those taken in regions of local maxima
in the K-band light, appear to have an excess continuum source, which
we interpret as hot dust with an effective temperature of $\sim$1000K.
The simplest way to confirm this conclusion would be to combine
spectroscopy of the type presented here with HKL imaging of the same
regions, which should reveal strong local excesses in the L-band
light, and with spatially-resolved metallicity measurements.

\begin{acknowledgements}

We thank the anonymous referee for many useful comments which
significantly improved the content and presentation of the paper. The
United Kingdom Infrared Telescope is operated by the Joint Astronomy
Centre on behalf of the U.K. Particle Physics and Astronomy Research
Council.

\end{acknowledgements}

\end{document}